\documentclass[conference, 9pt]{IEEEtran}
\IEEEoverridecommandlockouts
\usepackage{url}
\usepackage{cite}
\usepackage{amsmath,amssymb,amsfonts}
\usepackage{graphicx}
\usepackage{textcomp}
\usepackage{xcolor}

\DeclareMathOperator*{\argmin}{argmin}
\usepackage{graphicx,cite,amsmath,amssymb,amsfonts,amsthm,mathtools,algorithm,mathabx,textcomp,blindtext,xcolor,float,mwe,dblfloatfix,bm, xfrac,algpseudocode, stackengine, verbatim, subcaption}
\usepackage{enumitem}
\usepackage[font=scriptsize]{subcaption}

\algtext*{EndIf}
\algtext*{EndFor}

\def\BibTeX{{\rm B\kern-.05em{\sc i\kern-.025em b}\kern-.08em
    T\kern-.1667em\lower.7ex\hbox{E}\kern-.125emX}}
\begin{document}

\title{A Multi-Agent Multi-Environment Mixed Q-Learning for Partially Decentralized Wireless Network Optimization \vspace{-5pt} \\
\thanks{This work was funded by the following grants: NSF CCF-1817200, ARMY W911NF1910269, ARO W911NF2410094, DOE DE-SC0021417, Swedish Research Council 2018-04359, NSF CCF-2008927, NSF CCF-2200221, NSF RINGS-2148313, UCD/NSF CCF-2311653, NSF A22-2666-S003, ONR 503400-78050, ONR N00014-15-1-2550, ONR N00014-22-1-2363, UCSD/NSF KR 705319, UCD/NSF A25-0139-S009.}}

\author{\IEEEauthorblockN{Talha Bozkus}
\IEEEauthorblockA{\textit{Department of Electrical and Computer Engineering} \\
\textit{University of Southern California, USA\vspace{-10pt}}}
\and
\IEEEauthorblockN{Urbashi Mitra}
\IEEEauthorblockA{\textit{Department of Electrical and Computer Engineering} \\
\textit{University of Southern California, USA\vspace{-10pt}}}
}

\maketitle
\newtheorem{remark}{Remark}
\newtheorem{proposition}{Proposition}
\newtheorem{corollary}{Corollary}

\begin{abstract}
$Q$-learning is a powerful tool for network control and policy optimization in wireless networks, but it struggles with large state spaces. Recent advancements, like multi-environment mixed $Q$-learning (MEMQ), improves performance and reduces complexity by integrating multiple $Q$-learning algorithms across multiple related environments so-called {\em digital cousins}. However, MEMQ is designed for centralized single-agent networks and is not suitable for decentralized or multi-agent networks. To address this challenge, we propose a novel multi-agent MEMQ algorithm for partially decentralized wireless networks with multiple mobile transmitters (TXs) and base stations (BSs), where TXs do not have access to each other's states and actions. In uncoordinated states, TXs act independently to minimize their individual costs. In coordinated states, TXs use a Bayesian approach to estimate the joint state based on local observations and share limited information with leader TX to minimize joint cost. The cost of information sharing scales linearly with the number of TXs and is independent of the joint state-action space size. The proposed scheme is 50\% faster than centralized MEMQ with only a 20\% increase in average policy error (APE) and is 25\% faster than several advanced decentralized $Q$-learning algorithms with 40\% less APE. The convergence of the algorithm is also demonstrated.
\end{abstract}


\section{Introduction}
$Q$-learning is a powerful tool for optimization in wireless networks that can be modeled as Markov Decision Processes (MDPs) \cite{q_learning_wireless_2, q_learning_wireless_3, q_learning_wireless_4} but it faces challenges such as high estimation bias and variance, training instability, high sample complexity and slow convergence in large MDPs \cite{q_learning_survey, double_q, speedy_q, maxmin_q, averaged_dqn}. MEMQ has been proposed to address these issues by using multiple $Q$-learning algorithms across multiple, distinct yet related synthetic environments, or \emph{digital cousins} and fusing their $Q$-functions into a single ensemble estimate using adaptive weighting. MEMQ combines real-time sampling with synthetic environments running in parallel, and accelerates exploration and training compared to several state-of-the-art $Q$-learning and ensemble model-free reinforcement learning (RL) algorithms \cite{talha_eusipco, talha_icassp, pn_journal, ln_journal, talha_asilomar}. 

MEMQ was originally designed for single-agent centralized wireless networks to solve real-world challenges such as resource allocation, interference minimization and collision avoidance in MISO and MIMO networks with lower complexity and strong theoretical guarantees. However, it is not directly applicable to multi-agent or decentralized wireless networks. To this end, multi-agent $Q$-learning theory offers some tools to extend single-agent MEMQ to multi-agent networks \cite{sparse_cooperative, Hysteretic_q, independent_joint_q_3, coordinated_State_1, coordinated_State_2, coordinated_State_3, coordinated_State_4, coordinated_State_5}. \textit{Independent Learners (IL)} is a decentralized approach, where each agent independently learns its $Q$-functions, ignoring other agents' actions. This method is straightforward but may not guarantee convergence. \textit{Joint Learners (JL)} is a centralized approach and treats all agents as a single entity, directly learning joint actions through communication between agents. This approach yields optimal joint actions, yet it can be costly and inefficient with large number of agents. \textit{Mixed Learners (ML)} is a partially decentralized approach with information sharing, where the state-space is divided into coordinated and uncoordinated states. Agents independently update individual $Q$-tables in uncoordinated states and coordinate by sharing information with each other in coordinated states to update joint $Q$-table, which can be stored by one of the agents. ML leverages the performance of JL and the simplicity of IL. In this paper, we adopt a strategy similar to Mixed Learners with two improvements compared to previous work \cite{sparse_cooperative, Hysteretic_q, independent_joint_q_3, coordinated_State_1, coordinated_State_2, coordinated_State_3, coordinated_State_4, coordinated_State_5}, (i) we apply ML approach to MEMQ algorithms, which offer faster convergence than standard $Q$-learning. (ii) In coordinated states, agents estimate the joint state using a Bayesian approach based on their local observations in a decentralized way, minimizing the cost of information exchange.

The main contributions of this paper are as follows: (i) We propose a novel multi-agent partially decentralized MEMQ algorithm. (ii) We propose a Bayesian approach where agents estimate the joint state based on local aggregated received signal strength observations while interacting with leader TX by sharing minimal information to minimize joint costs only when needed for coordination. At other times, agents behave independently. The cost of information-sharing scales linearly with the number of TXs and is independent of the joint state-space size. (iii) We simulate our algorithm in a multi-agent partially decentralized wireless network with multiple mobile transmitters and base stations, capturing key aspects of real-world networks such as dimensionality, complexity, and privacy. Our algorithm is 50\% faster than centralized MEMQ with only 20\% more APE and is 25\% faster than several advanced decentralized $Q$-learning algorithms while reducing their APE by 40\%. We also show the fast convergence of our algorithm and the accuracy of the state estimation algorithm.

Vectors are bold lower case (\textbf{x}); matrices are bold upper case (\textbf{A}); sets are in calligraphic font ({$\mathcal{S}$}); and scalars are non-bold ($\alpha$).

\section{System Model and Tools}
\label{sec:system_model}

\subsection{Markov Decision Processes}
\label{subsec:mdp}
A \textbf{single agent MDP} is characterized by a 4-tuple $\{\mathcal{S}_i, \mathcal{A}_i, p_i, c_i\}$, where $\mathcal{S}_i$ and $\mathcal{A}_i$ denote the state and action spaces of agent $i$. The state and action at time $t$ are $s_{i,t}$ and $a_{i,t}$, respectively. The transition probability from $s_{i,t}$ to $s_{i,t}'$ under action $a_{i,t}$ is $p_{a_{i,t}}(s_{i,t}, s_{i,t}')$, and the incurred cost is $c_i(s_{i,t}, a_{i,t})$. The agent aims to solve the following infinite-horizon discounted cost minimization \cite{bertsekas_book}:
\vspace{-3pt}
\begin{align}\label{Equ:single_agent_minimization}
    v_i^{*}(s_i) = \min_{\bm{\pi}_i} \mathbb{E}_{\bm{\pi}_i} \sum_{t=0}^{\infty} \gamma^t c_i(s_{i,t}, a_{i,t}),
    \vspace{-5pt}
\end{align} 
where $v_i^{*}$ is the optimal value function of agent $i$, and $\gamma \in (0,1)$ is the discount factor. A \textbf{multi-agent MDP} generalizes a single-agent MDP to $N$ agents, each with its own states and actions. The joint state at time $t$ is $\bar{s}_t = (s_{1,t}, \ldots, s_{N,t})$ and the joint action is $\bar{a}_t = (a_{1,t}, \ldots, a_{N,t})$. The joint state and action spaces are $\mathcal{S} = \otimes_{i=1}^N \mathcal{S}_i$ and $\mathcal{A} = \otimes_{i=1}^N \mathcal{A}_i$ ($\otimes$: Kronecker product). Transition from joint state $\bar{s}_t$ to $\bar{s}'_t$ under joint action $\bar{a}_t$ occurs with probability $p(\bar{s}_t, \bar{a}_t, \bar{s}'_t)$, and stored in the $\{s,s',a\}^{th}$ element of the joint probability transition tensor (PTT) $\mathbf{\bar{P}}$ and incurs joint cost $c_i(\bar{s}_t, \bar{a}_t)$ for agent $i$. The multi-agent solves the following to find optimal joint value function $\bar{v}^{*}$ :\vspace{-5pt}
\begin{align}\label{Equ:multi_agent_minimization}
    \bar{v}^{*}(\bar{s}) = \min_{\bar{\pi}} \mathbb{E}_{\bar{\pi}} \sum_{t=0}^{\infty} \gamma^t \sum_{i=1}^N c_i(\bar{s}_{t}, \bar{a}_{t}).
\end{align}
\vspace{-25pt}
\subsection{$Q$-Learning}
A \textbf{single-agent $Q$-learning} solves (\ref{Equ:single_agent_minimization}) by learning $Q$-functions of agent $i$ \(Q_i(s_i, a_i)\)  and storing them in a $Q$-table using rule \cite{bertsekas_book}:
\begin{equation}\label{Equ: Q-learning-update-rule}
    \hspace{-4pt}Q_i(s_i, a_i) \text{ = } (1\text{-}\alpha) Q_i(s_i, a_i) \text{+} \alpha (c_i(s_i, a_i) \text{+} \gamma \min_{a_i' \in \mathcal{A}_i} Q_i(s_i', a_i')),
\end{equation}
where $\alpha \in (0,1)$ is the learning rate. The $\epsilon$-greedy policies balance exploration and exploitation. $Q$-functions converge to their optimal values with probability one, given certain conditions \cite{q_learning_convergence}. The optimal policy is $\bm{\pi}_i^*(s_i) = \argmin_{a_i \in \mathcal{A}_i} Q_i^*(s_i, a_i)$, and the relationship between the optimal value function and optimal $Q$-functions is $v_i^*(s_i) = \min_{a_i \in \mathcal{A}_i} Q_i^*(s_i, a_i)$. As discussed previously, one can use independent learners (IL), joint learners (JL), or mixed learners (ML) to extend single-agent $Q$-learning to \textbf{multi-agent $Q$-learning} to solve (\ref{Equ:multi_agent_minimization}) and obtain $\bar{v}^*(\bar{s}) = \min_{\bar{a} \in \mathcal{A}} \bar{Q}^*(\bar{s},\bar{a})$ where $\bar{Q}^*$ is the optimal joint $Q$-functions.

\subsection{Multi-environment Mixed $Q$-Learning (MEMQ)}\label{Subsec:memq}

MEMQ leverages the concept of \textit{digital cousins} \cite{ln_journal}: the agent interacts with $K$ structurally related, but distinct environments, one real environment (Env$_1$) and $K$-1 synthetic environments Env$_n$ (digital cousins of the real environment), where $n$$>$2 is the environment order -- the power used to create the PTT of the synthetic environment.

The agent first estimates PTT of the real environment $\mathbf{P}$ by interacting with the real environment in real-time and collecting samples efficiently \cite{talha_eusipco}. Let $\mathbf{\hat{P}}_i$ denote the estimated PTT by agent $i$. Simultaneously, the agent constructs the PTT of the synthetic environments by taking the $n$-th power of the estimated PTT of the real environment $\mathbf{\hat{P}}^n_i$. Thus, $\mathbf{\hat{P}}^n_i$ contains the $n$-hop state-action transition information. Then the agent runs $K$ independent $Q$-learning algorithms across $K$ environments, and produce $Q$-functions, denoted as $Q^{(n)}_i$ for environment $\text{Env}_n$. These $Q$-functions are combined into a single ensemble $Q$-function, $Q^e_i$ using the following update rule:\vspace{-2pt} 
\begin{align}
    Q_i^e \gets uQ_i^e+(1-u)\sum_{n}w_i^{(n)}Q_i^{(n)}
\end{align}
where $w_i^{(n)}$ is the weight of Env$_n$ of TX$_i$ (inversely proportional to $||Q_i^{(n)}-Q_i^{(1)}||_2$) and $u$ is the update ratio. It is proven that $Q^e_i$ converges to the optimal $Q$-functions (in mean-square) faster than $Q^{(n)}_i$ \cite{pn_journal, talha_github_Pn_paper}. The sizes of $Q$-tables for $Q^{(n)}_i$ and $Q^e_i$ are $|\mathcal{S}_i| \times |\mathcal{A}_i|$.

Herein, $K$ and $n$ are hyperparameters. While $K$ is usually chosen small for modest-sized state-action spaces, selecting $n$ is more complex. For $K=4$, there are many possible sets of orders (1,2,3,4 or 1,2,3,5 or similar). The goal is to use the fewest synthetic environments that provide the most diverse multi-time-scale information. To achieve this, we employ the coverage-coefficient selection algorithm from \cite{talha_spawc, talha_coverage_journal}, which selects the synthetic environments based on their sample coverage of the real environment. 
\vspace{-6pt}
\subsection{Multi-Agent Partially Decentralized Wireless Network}
\begin{figure}[t]
    \scriptsize
    \centering
    \includegraphics[width=0.23\textwidth]{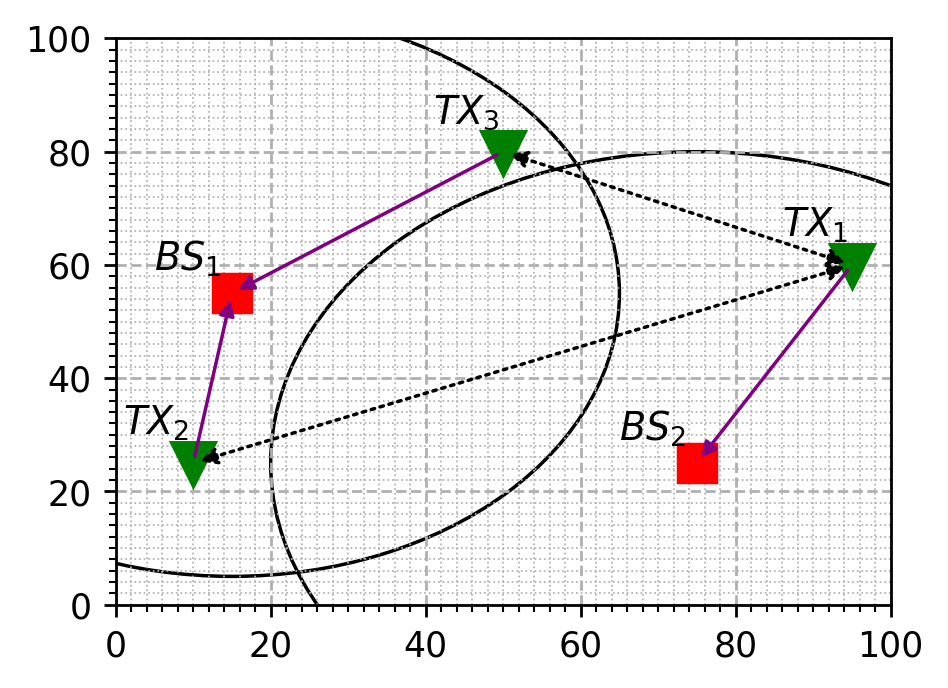}
    \vspace*{-0.05in}
    \caption{Example wireless network setting}
    \label{Fig:eaxmple_wireless_network}
    \vspace{-18pt}
\end{figure}

While our approach is general and has wide applications, to make our method and associated analysis accessible, we first describe a particular application framework. We consider a grid environment of size $L \times L$ with cells of size $\Delta_{L} \times \Delta_{L}$. There are $N_T$ mobile transmitters (TXs) and $N_B$ fixed base stations (BSs), where each TX can move randomly in four directions (U, R, D, L) or stay stationary, all with equal probability. TXs always remain on grid points. Each BS$_i$ has a circular coverage area with radius $r_i$; a TX can connect to BS only on or within this area. Fig.\ref{Fig:eaxmple_wireless_network} shows an example: red squares for BSs, green triangles for TXs, black solid circles for coverage areas, purple solid arrows for TX-BS association, and black dotted arrows for the communications between TXs. TX$_1$ is the leader TX with the largest coverage area ($r_i$), as it can communicate with and broadcast to the largest portion of the network. TXs can exchange their coverage area information to identify this leader. On the other hand, each TX can continuously measure (sense) the aggregated received signal strength (ARSS) due to all other TXs.

The TXs are the agents; hence, there are $N_T$ agents. The state of TX$_i$ is $s_i = (x_i, y_i, I_i)$: $(x_i, y_i)$ is its location with $x_i, y_i \in [0, L]$, and $I_i$ is ARSS. Since ARSS is continuous, we quantize \(I_i\) to the nearest value in \([I_{\text{min}}, I_{\text{max}}]\), which is discretized into \(\Delta_I\) logarithmically spaced bins. Agents do not have access to each others' states but estimate them based on local ARSS observations (see next section). The joint state $\bar{s}$ is \textbf{coordinated} if there is at least one agent $i$ such that $I_{i} > I_{thr}$, where $I_{thr}$ is the predefined ARSS threshold; otherwise, it is \textbf{uncoordinated}. Hence, if at least one TX decides that the state is coordinated, they update TX$_1$, which broadcasts this information to all other TXs. Let $|\mathcal{S}_C|$ and $|\mathcal{S}_U|$ be the number of coordinated and uncoordinated states. The action set of TX$_i$ is $\mathcal{A}_i$=$\{1, 2, \ldots, N_B\}$, where each index corresponds to a BS. If a TX is outside a BS's coverage area, the action is invalid and assigned an infinite cost to prevent selection. The cost function for TX$_i$ is the convex combination of the following three components:

\textit{1-Efficiency cost} = $\frac{P_{i}}{\log_2(1 + \operatorname{SNR}_i)}$, $\operatorname{SNR}_i = \frac{P_{i}}{d_i^2(n + I_i)}$, $d_i$ is distance between TX$_i$ and BS it is connected to, $P_{i}$ is the transmission power of TX$_i$ and $n \sim \mathcal{N}(0,\sigma^2)$. A higher SNR with constant transmitting power improves the effective data rate and reduces cost.

\textit{2-Reliability Cost} = $Q(k \sqrt{\text{SNR}_i})$,
where $k$ depends on the modulation scheme. For BPSK and QPSK, the $Q$-function represents the inverse relationship between SNR and Bit Error Rate (BER). Thus, higher SNR reduces BER and improves transmission reliability.

\textit{3-Fairness cost:} $Q\left(\frac{I_{thr}-I_{i}}{\sigma}\right)$. If the measured local ARSS exceeds the threshold, it suggests that some TXs are close to each other and may compete for the same resources (time slots, frequencies), potentially overloading some BSs while underutilizing others.

This wireless network is \textbf{partially decentralized}: (i) the cost function of each TX relies solely on local ARSS observations. (ii) TXs use only local observations data to estimate the joint state (see next section). (iii) TXs only interact with TX$_1$ by sharing limited information to minimize joint cost in coordinated states, preventing constant communication between all TXs. (iv) Based on experiments, $|\mathcal{S}_C| \ll |\mathcal{S}_U|$, which allows TXs to operate mostly independently, reducing communication and coordination costs (see next section).
\vspace{-2pt}
\section{Algorithm}
\label{sec:algorithm}

\begin{figure}[t]
    \centering
    \subfloat[\scriptsize Standard Q]{{\includegraphics[width=1.6cm]{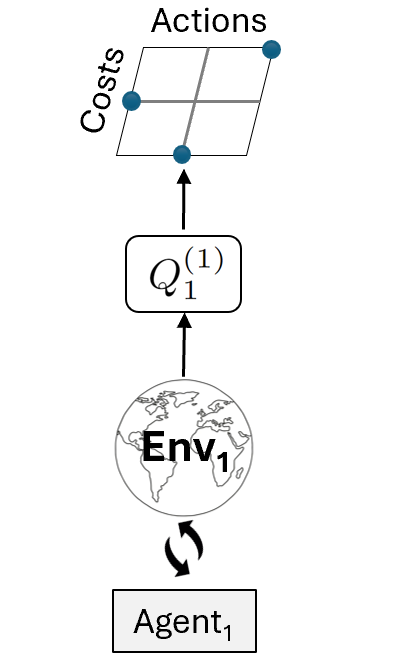}}}%
    \hspace{-6pt}
    \subfloat[\scriptsize MEMQ]{{\includegraphics[width=1.5cm]{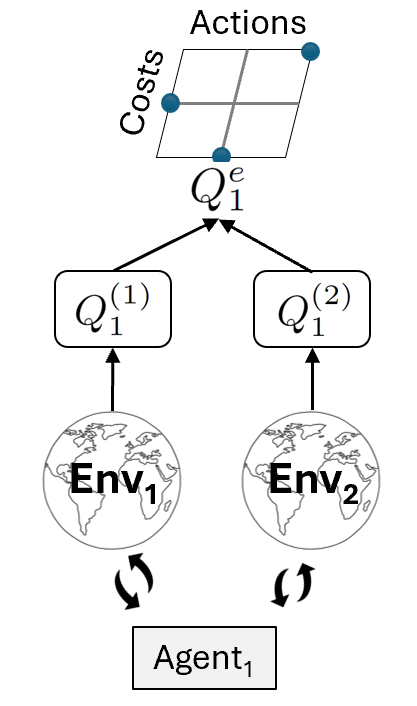}}}%
    \hspace{-1pt}
    \subfloat[\scriptsize Proposed Multi-Agent MEMQ]{{\includegraphics[width=4.7cm]{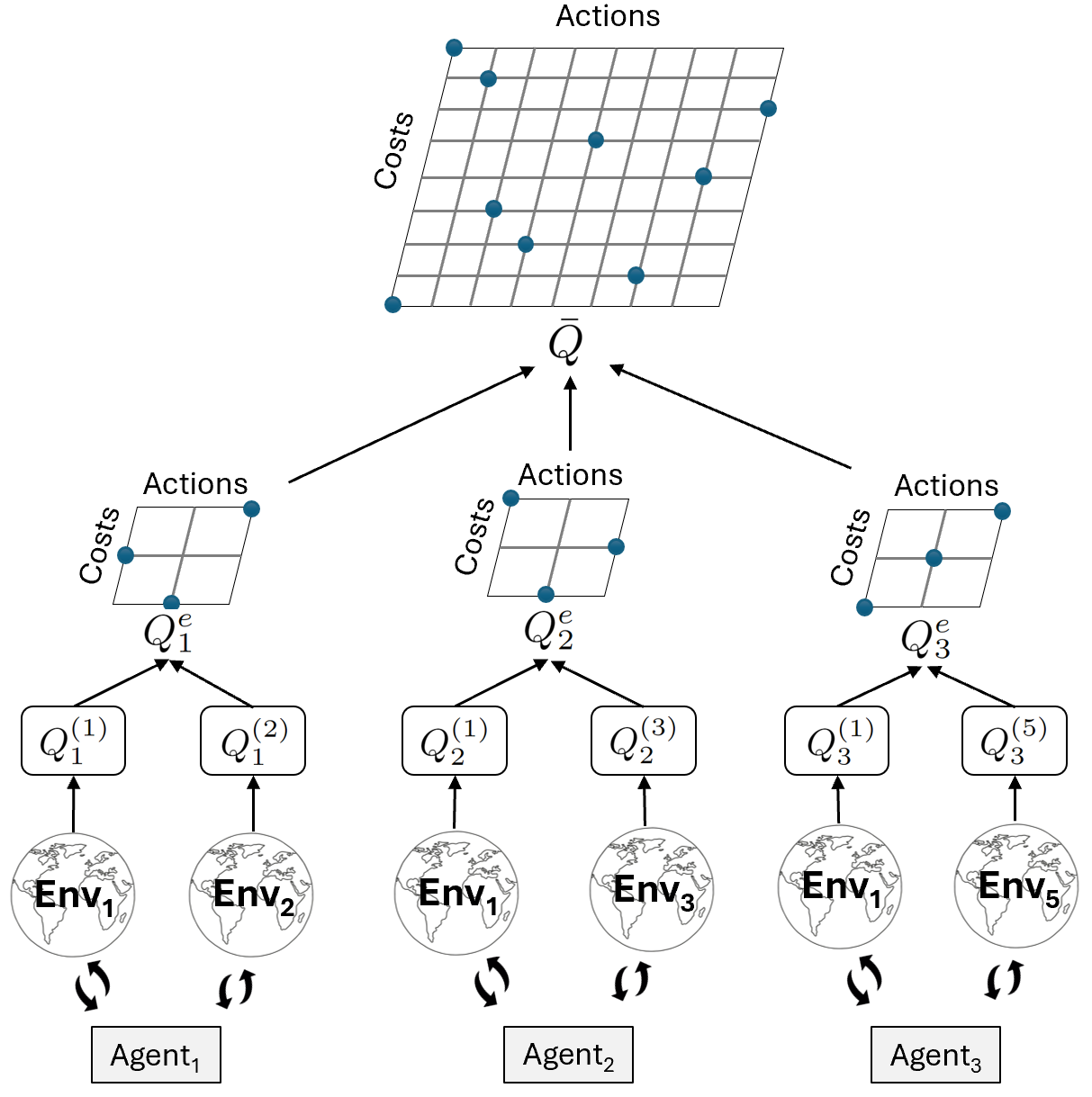}}}
    \caption{Comparison of Q-learning algorithms}
    \vspace{-2pt}
    \label{Fig:different_QL_algorithms}
\end{figure}

We describe our algorithm (Algorithm 1) using a three-agent network ($N_T=3$) to explain key details. However, our approach works for an arbitrary number of agents. Our simulations also explicitly consider larger values of $N_T$. We use the same notation in Section \ref{Subsec:memq}: $Q^{(n)}_i$ is the local $Q$-function for Env$_n$ and $Q^e_i$ is local ensemble $Q$-table of agent $i$ for $i=1,2,3$. There is also a single joint $Q$-table $\bar{Q}$, stored at TX$_1$, the leader TX. Different $Q$-functions are summarized in Table I. We use $Q$-function and $Q$-table notations interchangeably. Fig.\ref{Fig:different_QL_algorithms} compares standard $Q$-learning, standard MEMQ, and the proposed multi-agent MEMQ.

\setlength{\textfloatsep}{5pt}

\textbf{(1) -- Initialization:} Our algorithm uses the following inputs: trajectory length $l$, discount factor $\gamma$, learning rate $\alpha_t$, exploration probability $\epsilon_t$, update ratio $u_t$, initial belief vectors $\mathbf{b}_{i,0}$ for TX$_i$ at $t=0$, number of iterations $T$, and wireless network parameters. We assume each TX interacts with 2 environments: the real environment (Env$_1$) and a single synthetic environment (digital cousin), as shown in Fig.\ref{Fig:different_QL_algorithms}. The order of the synthetic environments for three TXs are 2, 3, and 5, respectively, and are chosen as in Sec.\ref{Subsec:memq}. Hence, each TX keeps three different local $Q$-tables: TX$_1$: $Q_1^{(1)}$, $Q_1^{(2)}$, $Q_1^{e}$. TX$_2$: $Q_2^{(1)}$, $Q_2^{(3)}$, $Q_2^{e}$. TX$_3$: $Q_3^{(1)}$, $Q_3^{(5)}$, $Q_3^{e}$ as in Table I. These $Q$-tables are \textbf{local} and \textbf{not shared} with other TXs. The joint $Q$-table $\bar{Q}(\bar{s}, \bar{a})$ is stored at TX$_1$ and of size $|\mathcal{S}| \times |\mathcal{A}|$. TX$_1$ updates this joint $Q$-table by interacting with other TXs \textbf{only} when coordination is necessary. Outputs are the joint $Q$-table and joint policy. TXs and BSs are also located randomly within the grid, ensuring no collisions. The belief vectors are initialized based on the initial local ARSS measurements as described in \cite{talha_github_icassp}. Herein, $K=2$ yields optimal performance for the given wireless network, yet our approach is compatible with any $K$; other wireless networks \cite{colink_journal, ln_journal} may require a larger $K$.

\textbf{(2) -- Sampling and Estimation:} Each TX$_i$ independently interacts with the real environment Env$_1$, collects samples \(\{s_i, a_i, s_i', c_i\}\), and stores this data in a shared buffer $\mathcal{B}$. Then, each TX$_i$ randomly samples from $\mathcal{B}$ to update their estimated PTT for Env$_1$ ($\mathbf{\hat{P}}_i$) by efficient sampling \cite{talha_eusipco}, and simultaneously builds and updates the PTT of its synthetic environment ($\mathbf{\hat{P}}_i^n$) as in Sec.\ref{Subsec:memq}. We reset belief updates and reinitialize belief vectors every $l$ steps to prevent error accumulation over time and improve adaptation to new observations.

\textbf{(3) -- Running independent MEMQ algorithms:} Each TX$_i$ runs an independent MEMQ algorithm using its two environments and continuously updates the ensemble $Q$-function $Q^{e}_i$. It is proven that \( Q^e_i \to Q^* \) for all $i$ if $u_t$ is chosen time-varying such that $u_t \xrightarrow{t \rightarrow \infty} 1$ \cite{pn_journal}. Although the same parameters are used in each MEMQ, the different underlying PTTs and the inherent randomness in sampling and exploration result in varying convergence rates for each MEMQ.

\begin{algorithm}[t]
\small
\caption{Multi-Agent Partially Decentralized MEMQ}
\hspace*{\algorithmicindent} \textbf{Inputs:} $l$,$\gamma$,$\alpha_t$,$\epsilon_t$,$u_t$,$\mathbf{b}_{i,0}$,$Q_i^{(n)}$,$Q^{e}_i$,wireless network parameters \\
\hspace*{\algorithmicindent} \textbf{Outputs:} $\bar{Q}, \bm{\bar{\pi}}$ (\textbf{Sec-1})
\begin{algorithmic}[1]
    \State Distribute TXs and BSs randomly (\textbf{Sec-1})
    \For{$t = 1$ to $T$}
        \For{each TX$_i$ \textbf{simultaneously}}
            \State Reinitialize $\mathbf{b}_{i,t}$ if $t$ $\%$ $l$ = 0 (\textbf{Sec-2})
            \State Interact with Env$_1$ independently, collect \(\{s_i, a_i, s_i', c_i\}\), and add to the common buffer $\mathcal{B}$ (\textbf{Sec-2})
            \State Sample \(\{s_i, a_i, s_i', c_i\}\) from $\mathcal{B}$ randomly to update $\mathbf{P}_i$ and the PTT of synthetic environment  (\textbf{Sec-2})
            \State Run MEMQ with two environments and inputs (\textbf{Sec-3})
            \State Decide if current state is coordinated based on local ARSS measurements. If so, estimate $\bar{s}_{i,t}, \bar{s}'_{i,t}$ individually
(\textbf{Sec-4})
            \State Update the $Q$-functions as in Table I (\textbf{Sec-5})
        \EndFor
    \EndFor
\State $\bar{\bm{\pi}}(\bar{s}) \gets \operatorname*{argmin}_{\bar{a}'}\bar{Q}(\bar{s},\bar{a}')$
\end{algorithmic}
\label{Algorithm: ensemble_link_learning}
\end{algorithm}

\begin{table*}[t]
\centering
\scriptsize
\setlength{\tabcolsep}{3pt}
\renewcommand{\arraystretch}{1.3}
\hspace{-90pt}
\begin{minipage}[t]{0.58\textwidth} % Adjust width as needed
\caption{\small Different Q-functions}\vspace{-5pt}
\centering
\begin{tabular}{|c|c|c|c|}
\hline
 & Individual & Ensemble & Joint \\ \hline
TX$_1$ & $Q^{(1)}_1, Q^{(2)}_1$ & $Q^e_1$ & $\bar{Q}$ \\ \hline
TX$_2$ & $Q^{(1)}_2, Q^{(3)}_2$ & $Q^e_2$ & - \\ \hline
TX$_3$ & $Q^{(1)}_3, Q^{(5)}_3$ & $Q^e_3$ & - \\ \hline
\end{tabular}
\end{minipage}
\hspace{-80pt} % Space between tables
\begin{minipage}[t]{0.65\textwidth} % Adjust width as needed
\renewcommand{\arraystretch}{1.3}
\caption{\small Q-function Update Rules for Different Cases}\vspace{-5pt}
\scriptsize
\centering
\begin{tabular}{|c|c|}
\hline
U $\rightarrow$ U &
$Q_{i}^{(n)}(s_i,a_i), Q^{e}_i(s_i,a_i) \leftarrow$ MEMQ\\
\hline
U $\rightarrow$ C &
$Q_{i}^{(n)}(s_i,a_i) \text{=} (1\text{-}\alpha)Q_{i}^{(n)}(s_i,a_i) \text{+} \alpha[c_i(s_i,a_i) \text{+}  \frac{\gamma}{3} \max_{\bar{a}'} \bar{Q}(\bar{s}', \bar{a}')]$, $Q^{e}_i(s_i,a_i) \leftarrow$ MEMQ\\
\hline
C $\rightarrow$ U &
$\bar{Q}(\bar{s},\bar{a}) = (1-\alpha)\bar{Q}(\bar{s},\bar{a}) + \alpha \sum_{i=1}^3[c_i(\bar{s},\bar{a}) + \gamma \max_{a_i'} Q_{i}^{e}(s_i', a_i')]$\\
\hline
C $\rightarrow$ C &
$\bar{Q}(\bar{s},\bar{a}) = (1-\alpha)\bar{Q}(\bar{s},\bar{a}) + \alpha[\sum_{i=1}^3c_i(\bar{s},\bar{a}) + \gamma \max_{\bar{a}'} \bar{Q}(\bar{s}', \bar{a}')]$\\
\hline
\end{tabular}
\end{minipage}
\vspace{-12pt}
\end{table*}

\textbf{(4) -- Estimating joint state using local ARSS measurements:} 
Each TX continuously measures ARSS due to all other TXs, which is crucial for managing coordination in coordinated states as ARSS depends on the distance between TXs, and hence reveals some information about the presence and proximity of other TXs. ARSS simulation details are provided in \cite{talha_github_icassp}. We describe the Bayesian approach for estimating the joint state, which is used only in the coordinated states. For TX$_1$, this involves estimating the joint state of TX$_2$ and TX$_3$ using the local ARSS measurement $I_{1,t}$ at time $t$:\vspace{-5pt}
\begin{align}
    \hat{s}_{2,t}, \hat{s}_{3,t} &= \underset{s_{2}, s_{3}}{\operatorname{argmax}} \hspace{4pt} p(s_{2}, s_{3} \mid I_{1,t}) \nonumber\\
    &= \underset{s_{2}, s_{3}}{\operatorname{argmax}} \hspace{4pt} p(I_{1,t} \mid s_{2}, s_{3}) \mathbf{b}_{1,t}(s_{2}, s_{3})\label{Equ: argmax}
    \vspace{-6pt}
\end{align}
where $p(I_{1,t} \mid s_{2}, s_{3})$ $\sim$ $ N(\bm{\mu}_{1,t}, \bm{\Sigma}_{1,t})$, and $\bm{\mu}_{1,t}$ and $\bm{\Sigma}_{1,t}$ are computed in real-time based on local ARSS measurements up to time $t$ (see \cite{talha_github_icassp} for computations). Herein, $\mathbf{b}_{i,t}$ denotes the TX$_i$'s belief vector at time $t$. The estimated joint state of the system by TX$_1$ is $\bar{s}_{1,t} = (s_{1,t}, \hat{s}_{2,t}, \hat{s}_{3,t})$. The belief for $(s_2,s_3)$ is updated as:\vspace{-5pt}
    \begin{equation}
        \mathbf{b}_{1,t+1}(s_{2}, s_{3}) \leftarrow f(I_{1,t} \mid s_{2}, s_{3}) \mathbf{b}_{1,t}(s_{2}, s_{3}),\label{Equ: argmax_belief}\vspace{-3pt}
    \end{equation}
which is then normalized such that the elements of $\mathbf{b}$ adds up to 1.

We emphasize that the argmax operation in (\ref{Equ: argmax}) is not an exhaustive search over the entire state space. The search at time $t$ is restricted to a subset of the state space \(\{s_{2,t}: \|s_{2,0} - s_{2,t}\|_2 \leq \Delta\} \times \{s_{3,t}: \|s_{3,0} - s_{3,t}\|_2 \leq \Delta\)\}, where $\Delta = \delta_0+2\Delta_L\operatorname{min}(l,t \% l)$ and the distance between the first estimated and correct states is at most $\delta_0$ (see \cite{talha_github_icassp}). Thus, the search space size is independent of the full state space size. We will numerically show that estimated states converge to the true states for small enough $\delta_0$ and modest values of $l, \Delta_L$. A small $\delta_0$ can be chosen by analyzing the historical behavior of TXs or leveraging domain knowledge. Without such knowledge, resetting the belief update every $l$ steps can prevent error accumulation over time. The joint state estimation for other TXs is done in a similar way; hence, each TX$_i$ at time $t$ has its own joint state estimate $\bar{s}_{i,t}$ with a belief vector $\mathbf{b}_{i,t}$. We will numerically show that knowledge of local ARSS is sufficient to produce good state estimates.

\textbf{(5) -- Updating $Q$-functions:} The $Q$-functions ($Q_i^{(n)}, Q^{e}_i, \bar{Q}$) are updated depending on the direction of transition between coordinated (C) and/or uncoordinated (U) states similar to \cite{sparse_cooperative}. Table 1 summarizes the rules. When U$\rightarrow$U, each TX independently updates their \(Q\)-functions as per Table 1. At the end, all TXs also share their \(Q\)-functions with TX\(_1\) once, and TX\(_1\) then updates the joint \(Q\)-function by summing the individual \(Q\)-functions: $\bar{Q}(\bar{s}, \bar{a}) = \sum_{i=1}^3 Q_i^e(s_i, a_i).$ In other cases, TXs estimate the joint state individually and share it with the corresponding belief probability with TX$_1$. TX$_1$ selects the joint state with the highest belief probability, broadcasts the corresponding joint action to all TXs, and each TX takes its individual action. TXs then share their individual costs and $Q$-functions with TX$_1$, which then updates the joint $Q$-function. This approach enables each TX to indirectly gain information from the other TXs via the leader TX, with a total information-sharing cost over $T$ iterations in Big-O notation as $O(N_T\max\{T, |\mathcal{S}_i||\mathcal{A}_i|\})$ (see \cite{talha_github_icassp} proof). The cost scales linearly with the number of TXs ($N_T$) and is independent of the joint state-action space size, unlike the centralized case.

If the belief vector in (\ref{Equ: argmax_belief}) has also a multivariate Gaussian form as $\mathbf{b}_{1,t} \sim N(\bm{\mu}'_{1,t}, \bm{\Sigma}'_{1,t})$, then it is easy to show that the uncertainty in the updated belief vector $\mathbf{b}_{1,t+1}$ monotonically decreases over time, i.e., $\bm{\Sigma}'_{1,t+1} \preceq \bm{\Sigma}'_{1,t}$. On the other hand, the updated mean depends on the local ARSS measurements, which vary as TXs move, and hence may not change monotonically over time.
\vspace{-4pt}
\section{Numerical Results}\label{sec:numerical_results}
\vspace{-4pt}

Let $\bm{\bar{\pi}^{*}}$ be the optimal joint policy minimizing (\ref{Equ:multi_agent_minimization}), $\bar{\bm{\pi}}$ be the estimated joint policy of Algorithm \ref{Algorithm: ensemble_link_learning}. We define the \textit{average policy error} as: APE = $\frac{1}{|\mathcal{S}|} \sum_{\bar{s}=1}^{|\mathcal{S}|} \mathbf{1}(\bm{\bar{\pi}^{*}}(\bar{s})$$\neq$$\bar{\bm{\pi}}(\bar{s}))$. We employ the following numerical parameters: $T$=5000, $l$=30, $\gamma$=0.95, $\alpha_t$=$\frac{1}{1+\sfrac{t}{1000}}$, $\epsilon_t$=max$(0.99^t, 0.01)$, $u_t$=1-$e^{\sfrac{-t}{1000}}$, $L$=100, $\Delta_L$=1, $r_i$$\sim$$\operatorname{unif}[\frac{L}{3},\frac{2L}{3}]$, $\Delta_I$=10, $\sigma$=$10^{-3}$, $k$=1, $P_i$=20dBm, $I_{max}$=10log($N_T$-1)-10dBm, $I_{thr}$=10log($N_T$-1)-45dBm, $I_{min}$=10log($N_T$-1)-60dBm. We employ the following algorithms for performance comparison: (i) Independent Q-learning (IQ) \cite{independent_q}  (ii) Hysteretic Q (HQ) \cite{Hysteretic_q}, (iii) Partaker-Sharer-Advising Q (PSAQ) \cite{partaker_q}, (iv) Sparse Cooperative Q (SCQ) \cite{sparse_cooperative}, and (v) MEMQ \cite{pn_journal}. Herein, (i), (ii), (iii) are decentralized, (iv) is partially decentralized, (v) is centralized. All algorithms employ multiple agents (MEMQ treats multiple agents as a single big agent). Simulations are averaged over 100 independent simulations. See \cite{talha_github_icassp} for details. 

\begin{figure}[t]
    \centering
    \subfloat[APE vs ($N_T, N_B$) \label{fig:ape_vsNbNt}]{{\includegraphics[width=4cm]{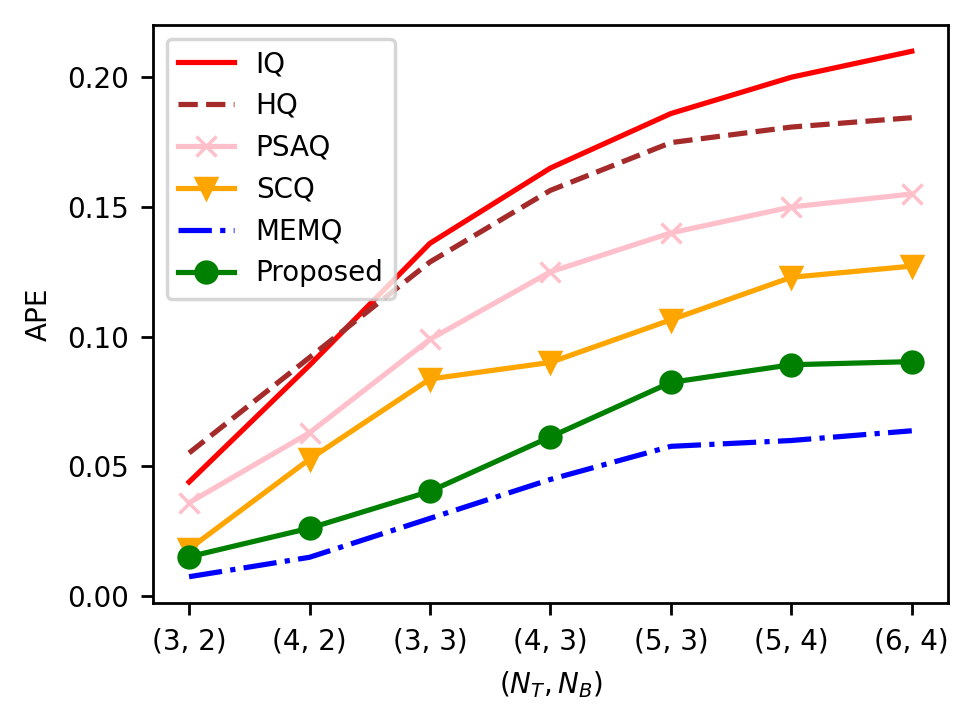}\vspace*{-0.12in}}}
    \subfloat[AQD vs iterations \label{fig:Qfunc_diff}]
    {{\includegraphics[width=4cm]{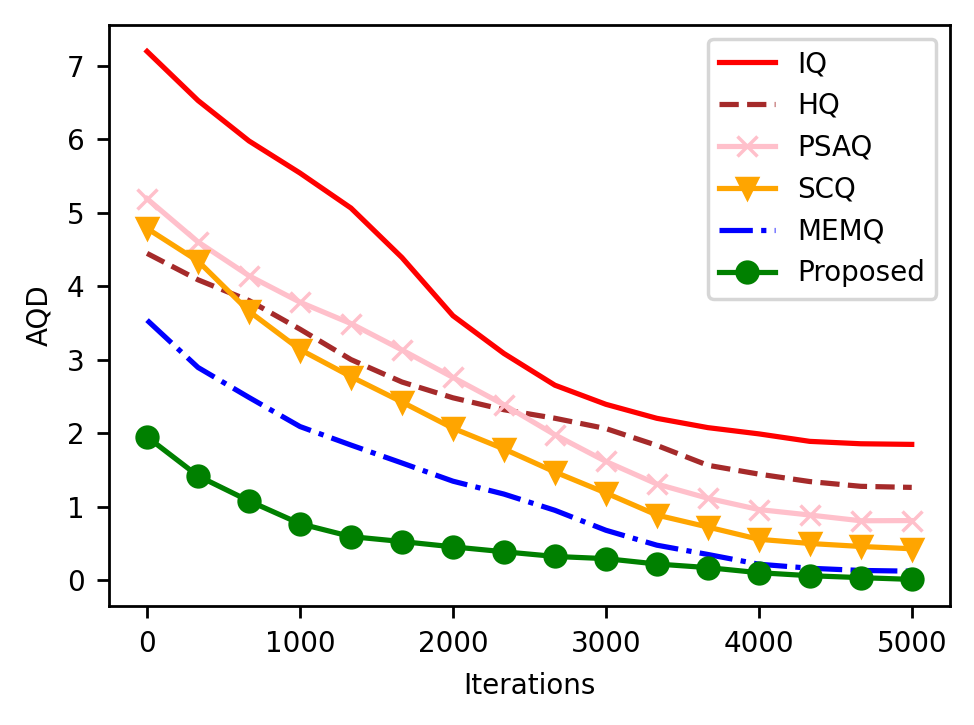}\vspace*{-0.12in}}}
    \vspace*{-0.03in}
    \subfloat[Runtime vs ($N_T, N_B$) \label{fig:runtime} ]{{\includegraphics[width=4cm]{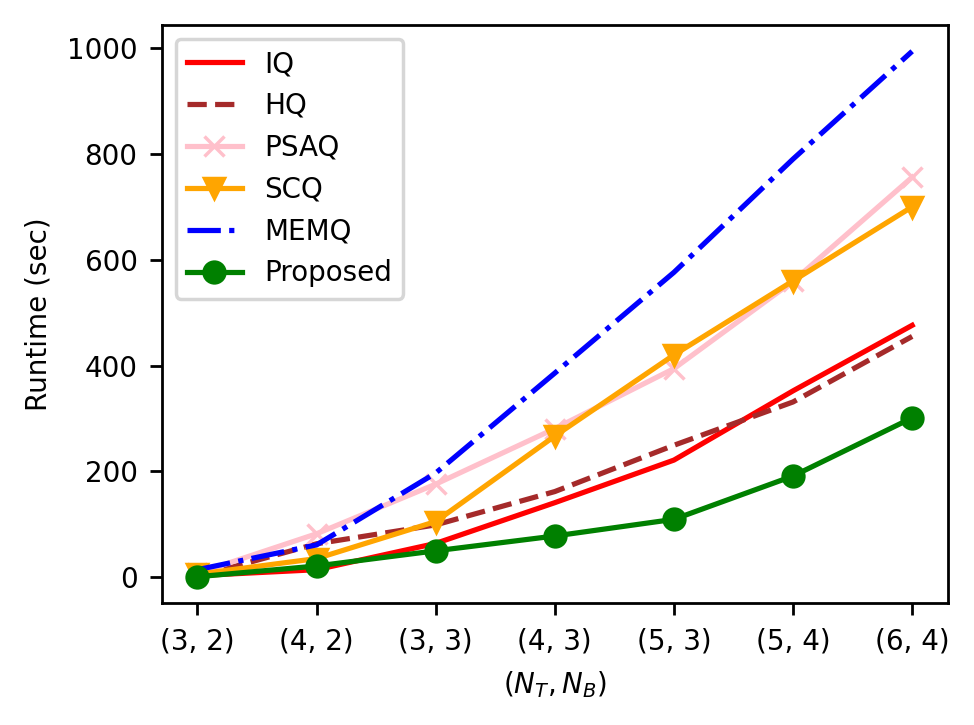}\vspace*{-0.08in}}}
    \subfloat[Estimated distance vs iterations  \label{fig:state_estimation} ]{{\includegraphics[width=4.4cm]{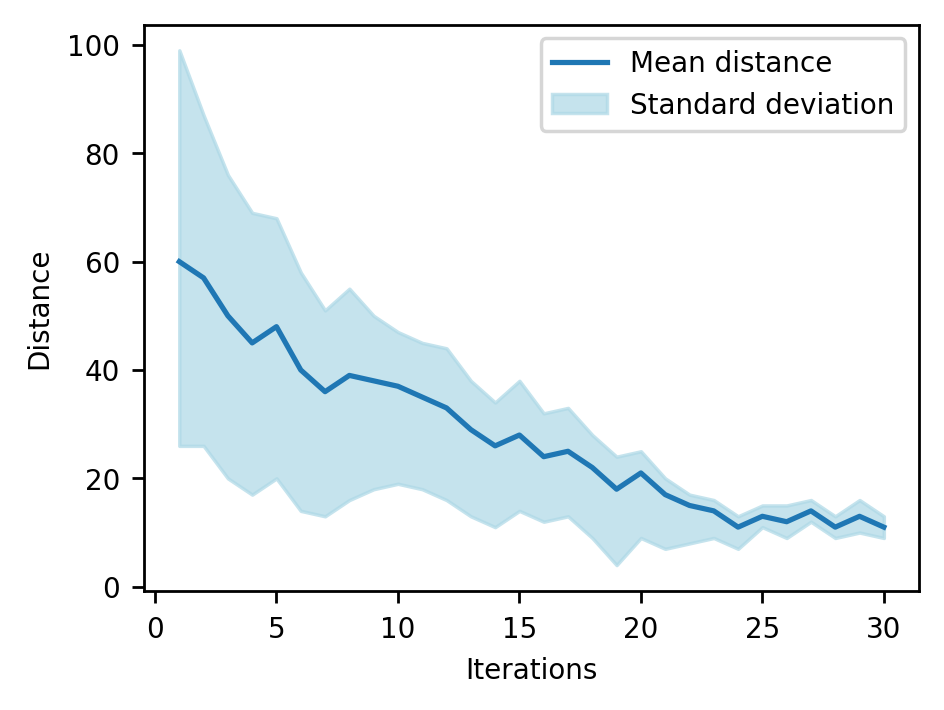}\vspace*{-0.08in}}}
    \vspace*{-0.05in}
    \caption{Numerical results}
    \vspace{-6pt}
\end{figure}

Fig.\ref{fig:ape_vsNbNt} shows the APE results of our proposed algorithm compared to various $Q$-learning algorithms across increasing $(N_T,N_B)$. Our algorithm has only 20\% higher APE than the centralized MEMQ, which achieves the lowest APE due to its access to global information. However, our algorithm outperforms others with 40\% less APE. The fully decentralized nature of IQ and HQ limits their performance compared to ours, PSAQ struggles with budget constraints on $Q$-function exchange, and SCQ lacks the ability to exploit multi-time-scale information from multiple environments.

Let $\bar{Q}^*$ be the optimal joint $Q$-functions. We define the \textit{average $Q$-function difference} as AQD = $\frac{1}{|\mathcal{S}||\mathcal{A}|} \sum_{\bar{s},\bar{a}} (\bar{Q}^{*}(\bar{s},\bar{a})-\bar{Q}(\bar{s},\bar{a}))^2$. Fig.\ref{fig:Qfunc_diff} shows the AQD results for various algorithms over iterations with $(N_T, N_B) = (5,3)$. Our algorithm converges to the optimal $Q$-functions 50\% faster than MEMQ, which suffers from an insufficient amount of exploration and samples for the joint system, whereas our algorithm efficiently leverages the purely decentralized nature of TXs in uncoordinated states. Other algorithms converge to specific values, but not necessarily to the optimal $Q$-functions, which is a common issue among many decentralized multi-agent algorithms \cite{independent_q, Hysteretic_q}.

Fig.\ref{fig:runtime} shows the runtime results for various algorithms across increasing $(N_T,N_B)$, reflecting computational complexity (time required for convergence with fine-tuned parameters on the same hardware). While all algorithms show increase in runtime as the network size grows, our algorithm has the smallest increase -- 25\% less than other algorithms and 50\% less than centralized MEMQ. This efficiency is due to (i) the purely decentralized approach in uncoordinated states, making it as fast as IQ and HQ, (ii) limited information exchange in coordinated states, similar to SCQ and PSAQ, (iii) fast exploration capabilities of digital cousins and fast convergence of standard MEMQ, (iv) the cost of information sharing which increases linearly with $N_T$. The percentage of coordinated states for different $(N_T,N_B)$ pairs are: 8\% for (3,2) and (3,3), 10\% for (4,2) and (4,3), 13\% for (5,3) and (5,4), and 16\% for (6,4). These three figures demonstrate that our algorithm offers the best performance-complexity trade-off, making it a practical and scalable choice for real-world wireless networks.

Fig.\ref{fig:state_estimation} shows the sum of $l_2$ distances between the true and estimated coordinates of TX$_2$ and that of TX$_3$ by TX$_1$ over $l = 30$ iterations using the state estimation approach with $(N_T, N_B) = (3,2)$. The blue solid curve represents the average distance over 100 simulations, while the blue-shaded area is the standard deviation. Although the initial estimate is far from optimal, TX$_1$ effectively utilizes its local ARSS observations to gradually improve its joint state estimate. In this example, TX$_1$ can estimate the joint location of TX$_1$ and TX$_2$ within $\approx$12m error. As TXs are likely to move during these 30 iterations, precise coordinate estimation is challenging. This suggests that knowledge of local ARSS is sufficient to capture interactions between different TXs, without requiring additional environment-specific metrics like channel state information.
\vspace{-4pt}
\section{Conclusions}\label{sec:conclusions}
\vspace{-4pt}
In this paper, we introduce a novel multi-agent MEMQ algorithm tailored to partially decentralized wireless networks, addressing the limitations of standard MEMQ and advanced $Q$-learning algorithms in large state-spaces. In uncoordinated states, each TX independently minimizes its cost function, while in coordinated states, TXs exchange limited information with leader TX to minimize the joint cost, and the cost of sharing scales linearly with the number of TXs. TXs use a Bayesian approach to estimate the joint state based on local ARSS observations. Our algorithm inherits key properties from standard MEMQ and converges to optimal $Q$-functions 50\% faster than centralized MEMQ with only a 20\% increase in APE and 25\% faster than several advanced $Q$-learning algorithms with 40\% less APE. Our future work focuses on establishing several theoretical guarantees on the complexity and convergence of our algorithm.

\bibliographystyle{IEEEtran}
\bibliography{references.bib}

\begin{thebibliography}{10}

\bibitem{q_learning_wireless_2}
Seungmin Lee, Heejung Yu, and Howon Lee,
\newblock ``Multiagent q-learning-based multi-uav wireless networks for maximizing energy efficiency: Deployment and power control strategy design,''
\newblock {\em IEEE Internet of Things Journal}, vol. 9, no. 9, pp. 6434--6442, 2021.

\bibitem{q_learning_wireless_3}
Dejan~V. Djonin and Vikram Krishnamurthy,
\newblock ``${Q}$-learning algorithms for constrained markov decision processes with randomized monotone policies: Application to mimo transmission control,''
\newblock {\em IEEE Transactions on Signal Processing}, vol. 55, no. 5, pp. 2170--2181, 2007.

\bibitem{q_learning_wireless_4}
Libin Liu and Urbashi Mitra,
\newblock ``On sampled reinforcement learning in wireless networks: Exploitation of policy structures,''
\newblock {\em IEEE Transactions on Communications}, vol. 68, no. 5, pp. 2823--2837, 2020.

\bibitem{q_learning_survey}
Jesse Clifton and Eric Laber,
\newblock ``Q-learning: Theory and applications,''
\newblock {\em Annual Review of Statistics and Its Application}, vol. 7, no. 1, pp. 279--301, 2020.

\bibitem{double_q}
Hado Hasselt,
\newblock ``Double q-learning,''
\newblock {\em Advances in neural information processing systems}, vol. 23, 2010.

\bibitem{speedy_q}
Mohammad Ghavamzadeh, Hilbert Kappen, Mohammad Azar, and R{\'e}mi Munos,
\newblock ``Speedy {Q}-learning,''
\newblock {\em Advances in neural information processing systems}, vol. 24, 2011.

\bibitem{maxmin_q}
Qingfeng Lan, Yangchen Pan, Alona Fyshe, and Martha White,
\newblock ``Maxmin {Q}-learning: Controlling the estimation bias of {Q}-learning,''
\newblock {\em CoRR}, vol. abs/2002.06487, 2020.

\bibitem{averaged_dqn}
Oron Anschel, Nir Baram, and Nahum Shimkin,
\newblock ``Averaged-dqn: Variance reduction and stabilization for deep reinforcement learning,''
\newblock in {\em International conference on machine learning}. PMLR, 2017, pp. 176--185.

\bibitem{talha_eusipco}
Talha Bozkus and Urbashi Mitra,
\newblock ``Ensemble link learning for large state space multiple access communications,''
\newblock in {\em 2022 30th European Signal Processing Conference (EUSIPCO)}, 2022, pp. 747--751.

\bibitem{talha_icassp}
Talha Bozkus and Urbashi Mitra,
\newblock ``Ensemble graph q-learning for large scale networks,''
\newblock in {\em ICASSP 2023 - 2023 IEEE International Conference on Acoustics, Speech and Signal Processing (ICASSP)}, 2023, pp. 1--5.

\bibitem{pn_journal}
Talha Bozkus and Urbashi Mitra,
\newblock ``Multi-timescale ensemble $q$-learning for markov decision process policy optimization,''
\newblock {\em IEEE Transactions on Signal Processing}, vol. 72, pp. 1427--1442, 2024.

\bibitem{ln_journal}
Talha Bozkus and Urbashi Mitra,
\newblock ``Leveraging digital cousins for ensemble q-learning in large-scale wireless networks,''
\newblock {\em IEEE Transactions on Signal Processing}, vol. 72, pp. 1114--1129, 2024.

\bibitem{talha_asilomar}
Talha Bozkus and Urbashi Mitra,
\newblock ``A novel ensemble q-learning algorithm for policy optimization in large-scale networks,''
\newblock in {\em 2023 57th Asilomar Conference on Signals, Systems, and Computers}. IEEE, 2023, pp. 1381--1386.

\bibitem{sparse_cooperative}
Jelle~R Kok and Nikos Vlassis,
\newblock ``Sparse cooperative q-learning,''
\newblock in {\em Proceedings of the twenty-first international conference on Machine learning}, 2004, p.~61.

\bibitem{Hysteretic_q}
La{\"e}titia Matignon, Guillaume~J Laurent, and Nadine Le~Fort-Piat,
\newblock ``Hysteretic q-learning: an algorithm for decentralized reinforcement learning in cooperative multi-agent teams,''
\newblock in {\em 2007 IEEE/RSJ International Conference on Intelligent Robots and Systems}. IEEE, 2007, pp. 64--69.

\bibitem{independent_joint_q_3}
Changxi Zhu, Ho-Fung Leung, Shuyue Hu, and Yi~Cai,
\newblock ``A q-values sharing framework for multi-agent reinforcement learning under budget constraint,''
\newblock {\em ACM Transactions on Autonomous and Adaptive Systems (TAAS)}, vol. 15, no. 2, pp. 1--28, 2021.

\bibitem{coordinated_State_1}
Yann-Micha{\"e}l De~Hauwere, Peter Vrancx, and Ann Now{\'e},
\newblock ``Learning multi-agent state space representations,''
\newblock in {\em Proceedings of the 9th International Conference on Autonomous Agents and Multiagent Systems: volume 1-Volume 1}, 2010, pp. 715--722.

\bibitem{coordinated_State_2}
Chao Yu, Minjie Zhang, and Fenghui Ren,
\newblock ``Coordinated learning for loosely coupled agents with sparse interactions,''
\newblock in {\em AI 2011: Advances in Artificial Intelligence: 24th Australasian Joint Conference, Perth, Australia, December 5-8, 2011. Proceedings 24}. Springer, 2011, pp. 392--401.

\bibitem{coordinated_State_3}
Francisco~S Melo and Manuela Veloso,
\newblock ``Learning of coordination: Exploiting sparse interactions in multiagent systems,''
\newblock in {\em Proceedings of The 8th International Conference on Autonomous Agents and Multiagent Systems-Volume 2}. Citeseer, 2009, pp. 773--780.

\bibitem{coordinated_State_4}
Yu-Jia Chen, Deng-Kai Chang, and Cheng Zhang,
\newblock ``Autonomous tracking using a swarm of uavs: A constrained multi-agent reinforcement learning approach,''
\newblock {\em IEEE Transactions on Vehicular Technology}, vol. 69, no. 11, pp. 13702--13717, 2020.

\bibitem{coordinated_State_5}
Yu-Jia Chen, Kai-Min Liao, Meng-Lin Ku, Fung~Po Tso, and Guan-Yi Chen,
\newblock ``Multi-agent reinforcement learning based 3d trajectory design in aerial-terrestrial wireless caching networks,''
\newblock {\em IEEE Transactions on Vehicular Technology}, vol. 70, no. 8, pp. 8201--8215, 2021.

\bibitem{bertsekas_book}
Dimitri Bertsekas,
\newblock {\em Reinforcement learning and optimal control},
\newblock Athena Scientific, 2019.

\bibitem{q_learning_convergence}
Francisco~S Melo,
\newblock ``Convergence of q-learning: A simple proof,''
\newblock {\em Institute Of Systems and Robotics, Tech. Rep}, pp. 1--4, 2001.

\bibitem{talha_github_Pn_paper}
Talha Bozkus and Urbashi Mitra,
\newblock ``Supplementary appendix,'' \url{https://github.com/talhabozkus/Multi-Timescale-Ensemble-QLearning}, 2023.

\bibitem{talha_spawc}
Talha Bozkus and Urbashi Mitra,
\newblock ``Coverage analysis of multi-environment q-learning algorithms for wireless network optimization,''
\newblock in {\em 2024 IEEE 25th International Workshop on Signal Processing Advances in Wireless Communications (SPAWC)}. IEEE, 2024, pp. 376--380.

\bibitem{talha_coverage_journal}
Talha Bozkus, Tara Javidi, and Urbashi Mitra,
\newblock ``Coverage analysis for digital cousin selection--improving multi-environment q-learning,''
\newblock {\em arXiv preprint arXiv:2411.08360}, 2024.

\bibitem{talha_github_icassp}
Talha Bozkus and Urbashi Mitra,
\newblock ``Supplementary appendix,'' \url{https://github.com/talhabozkus/icassp-25-appendix}, 2024.

\bibitem{colink_journal}
Talha Bozkus and Urbashi Mitra,
\newblock ``Link analysis for solving multiple-access mdps with large state spaces,''
\newblock {\em IEEE Transactions on Signal Processing}, vol. 71, pp. 947--962, 2023.

\bibitem{independent_q}
Caroline Claus and Craig Boutilier,
\newblock ``The dynamics of reinforcement learning in cooperative multiagent systems,''
\newblock {\em AAAI/IAAI}, vol. 1998, no. 746-752, pp. 2, 1998.

\bibitem{partaker_q}
Changxi Zhu, Ho-Fung Leung, Shuyue Hu, and Yi~Cai,
\newblock ``A q-values sharing framework for multi-agent reinforcement learning under budget constraint,''
\newblock {\em ACM Transactions on Autonomous and Adaptive Systems (TAAS)}, vol. 15, no. 2, pp. 1--28, 2021.

\end{thebibliography}

\end{document}